# ON THE INFLUENCE OF THE DATA SAMPLING INTERVAL ON COMPUTER-DERIVED K-INDICES

*A Bernard[1], M Menvielle[2,3], and A Chambodut[1*]*

[1*] Dept of Magnetic Observatories; Université de Strasbourg/EOST, CNRS; 5 rue Rene Descartes 67084 Strasbourg Cedex, France
Email: aude@unistra.fr
[2] Université Versailles St-Quentin; CNRS/INSU, LATMOS-IPSL, Guyancourt, France
Email: michel.menvielle@latmos.ipsl.fr
[3] Université de Paris-Sud, Département des Sciences de la Terre, Orsay, France

## ABSTRACT

*The K index was devised by Bartels et al. (1939) to provide an objective monitoring of irregular geomagnetic activity. The K index was then routinely used to monitor the magnetic activity at permanent magnetic observatories as well as at temporary stations. The increasing number of digital and sometimes unmanned observatories and the creation of INTERMAGNET put the question of computer production of K at the centre of the debate. Four algorithms were selected during the Vienna meeting (1991) and endorsed by IAGA for the computer production of K indices. We used one of them (FMI algorithm) to investigate the impact of the geomagnetic data sampling interval on computer produced K values through the comparison of the computer derived K values for the period 2009, January 1st to 2010, May 31st at the Port-aux-Français magnetic observatory using magnetic data series with different sampling rates (the smaller: 1 second; the larger: 1 minute). The impact is investigated on both 3-hour range values and K indices data series, as a function of the activity level for low and moderate geomagnetic activity.*

## 1  INTRODUCTION

The bias on *K* index values related to the sampling interval in digital recording of magnetic variations became of concern when magnetograms plotted from digital magnetometer data were put into use replacing analog magnetometers for K indices hand-scaling.

Niblett et al. (1984) used data from the Ottawa geomagnetic observatory to investigate the bias introduced by deriving the traditional K index from magnetograms plotted from recorded digital data. They hand-scaled *K* indices on analog photographic magnetograms ($K_A$) and on analog magnetograms reconstructed using digital magnetic data in the form of 1-minute averaged values or spot values at selected sampling intervals of 1-second or greater ($K_R$). They concluded that the lower *K* values tend to be biased significantly downward by one level when a digitizing interval greater than 30 seconds is used for construction of the reconstructed magnetograms.

In the course of the development and assessment of methods for computer derivation of K indices, differences between hand-scaled and computer derived K values were extensively studied, but the influence of the sampling interval was not addressed because at that time this sampling interval was always set to 1 minute (see, e.g., Menvielle et al., 1995 and reference therein; Bitterly et al., 1997). To our best knowledge, the influence of the sampling interval on the downward bias on the value of computer derived *K* indices has not yet been investigated. The present study investigates the impact of the change from 1-minute to 1-second in the INTERMAGNET standard sampling interval on computer derived *K* values, using Port-aux-Français magnetic observatory data for a 2-year period.

## 2  DERIVATION OF K INDICES FROM DIGITAL RECORDINGS

The *K* indices are based upon geomagnetic disturbances, measured in two horizontal geomagnetic components (*X* and *Y*), after eliminating the regular daily variation ($S_R$). An individual *K* index is an integer in the range 0 to 9 corresponding to a class that contains the largest range of geomagnetic disturbances in either of the two horizontal components during a 3-hour UT interval (Figure 1). Note that these limits may vary from one observatory to another since they depend on the corrected geomagnetic latitude of the observatory (Menvielle & Berthelier, 1991; Menvielle et al., 2011). In this study the two considered magnetic observatories, Ottawa (OTT) and Port-aux-Français (PAF), get a comparable absolute corrected geomagnetic latitude (OTT: 58.9°N; PAF: 58.8°S) and thus have the same classes of ranges (see Table 1).

| K indices | 0 | 1 | 2 | 3 | 4 | 5 | 6 | 7 | 8 | 9 |
|---|---|---|---|---|---|---|---|---|---|---|
| Range (nT) | 0 | 7.5 | 15 | 30 | 60 | 105 | 180 | 300 | 495 | 750  < 750 |

**Table 1.** Classes of ranges and corresponding K indices in OTT and PAF

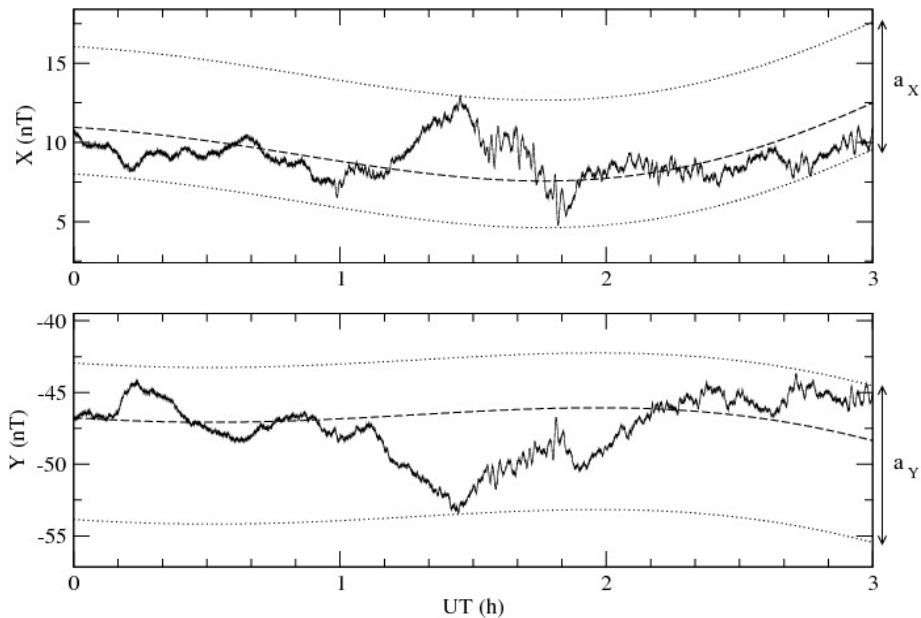

**Figure 1.** K index determination for the 0000-0300 UT three hour interval on 2010, April 11[th] at the PAF observatory. For both X and Y components, the dashed line represents the estimated $S_R$ variation, and the upper and lower dotted lines correspond to the $S_R$ curve shifted respectively up to the maximum and down to the minimum of the observed magnetic field deviation from the $S_R$ baseline; $a_X$ and $a_Y$ (respectively equal to 7.5 and 9.8 nT) correspond to the three hour ranges from which the K index is derived (K = 1 in this example).

The original definition of *K* indices (Bartels et al., 1939) requires hand scaling on analogue magnetograms. The question of the derivation of geomagnetic indices from digital data arose at the end of the seventies. Different algorithms enabling computer derivation of *K* indices were then developed and carefully assessed in the frame of an international comparison organised by the IAGA Working Group "Geomagnetic indices" (Coles & Menvielle, 1991; Menvielle, 1991). The reader is referred to Menvielle et al. (1995) for a complete review.

Basically, these algorithms estimate the $S_R$ variation from the magnetograms. The geomagnetic disturbances, from which the *K* indices are computed, are derived as the difference between the observed variations and the computer estimated $S_R$ ones. In the present study work, we use the so-called FMI algorithm (FMI for "Finnish Meteorological Institute") (Sucksdorff et al., 1991). Figure 2 shows an example of $S_R$ and *K* determination by the FMI algorithm for a moderately disturbed day (2010, April 12[th], 2010; *Am* = 30 nT) at the PAF observatory.

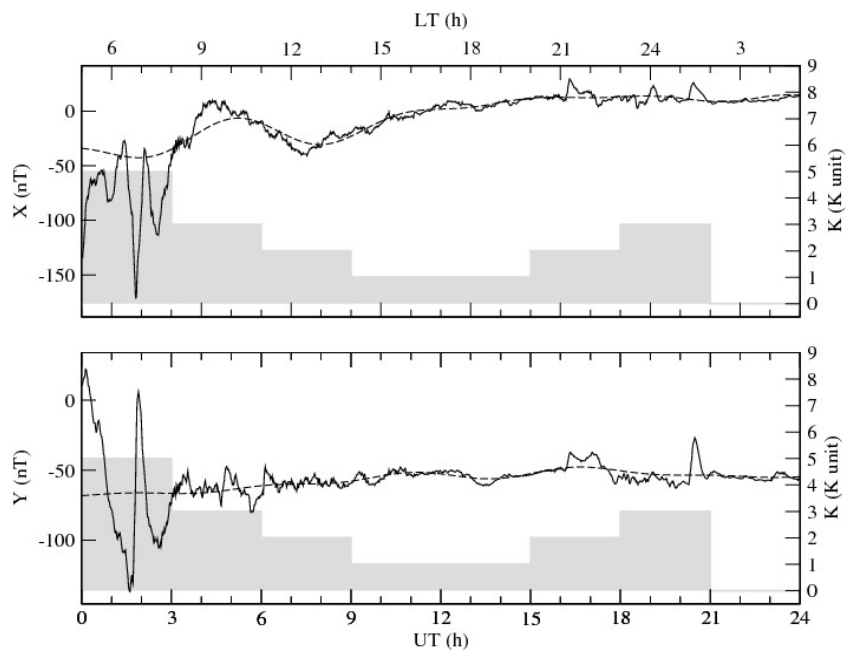

**Figure 2.** Magnetograms (minute values) of the horizontal components of the magnetic field recorded at the PAF observatory during a moderately disturbed magnetic day (April 12[th], 2010; Am = 30 nT). Variometer recordings: solid curves; $S_R$ : dashed curves; K indices: grey histograms.

## 3   DATA

The 1Hz data acquisition system at the PAF observatory was installed in March 2008. It consists of an internal digitization at a 10 Hz sampling rate of the variometer measurements. The obtained data series is then automatically filtered and rounded to the nearest second through an adapted INTERMAGNET Gaussian filter (Fotzé et al., 2007).

In the present work, we considered the 1-second data produced by the acquisition system and the classical 1-minute data time series computed from 1 Hz data strictly using the INTERMAGNET Gaussian filter as described in the INTERMAGNET Technical Reference Manual (2008). Two years (2008, July 14$^{th}$ to 2010, July 13$^{th}$) of 1-second recordings of the *X* (Geographic North) and *Y* (Geographic East) horizontal components are used. Our data set corresponds to a 2-year period during the deep solar minimum that occurred between solar cycles #23 and #24. During this period, the geomagnetic activity was very low, with a mean value of the *am* index equal to 8.6 nT.

The dataset used by Niblett et al. corresponds to a 1-year period (1982) just after the maximum of solar cycle #21, which was the second largest of the 20$^{th}$ century, from the point of view of the Wolf number. During this period, the geomagnetic activity was intense, with a mean value of the *am* index equal to 34.3 nT.

## 4   COMPARISON BETWEEN K FROM 1-SECOND AND 1-MINUTE VALUES

In the following, let $K_{60s}$ and $K_{1s}$ denote the *K* values derived using the 1-minute and 1-second sampling interval magnetograms, respectively.

$K_{60s}$ and $K_{1s}$ values corresponding to the same 3-hour interval might be different if the magnetic variations have energies in the frequency range between 1/2 and 1/120 Hz, which are the corresponding Nyquist frequencies respectively for the 1-second and 1-minute sampling rates.

Visual inspection of magnetograms shows that this is clearly the case during periods of geomagnetic activity. Figure 3 presents an example for which $a_X$ is underestimated when calculated from the 1-minute data time series. In this case, this leads to a one unit underestimation of the *X* component corresponding *K* value for the 0900-1200 interval: *K=2* (1-minute values) instead of *K=3* (1-second values).

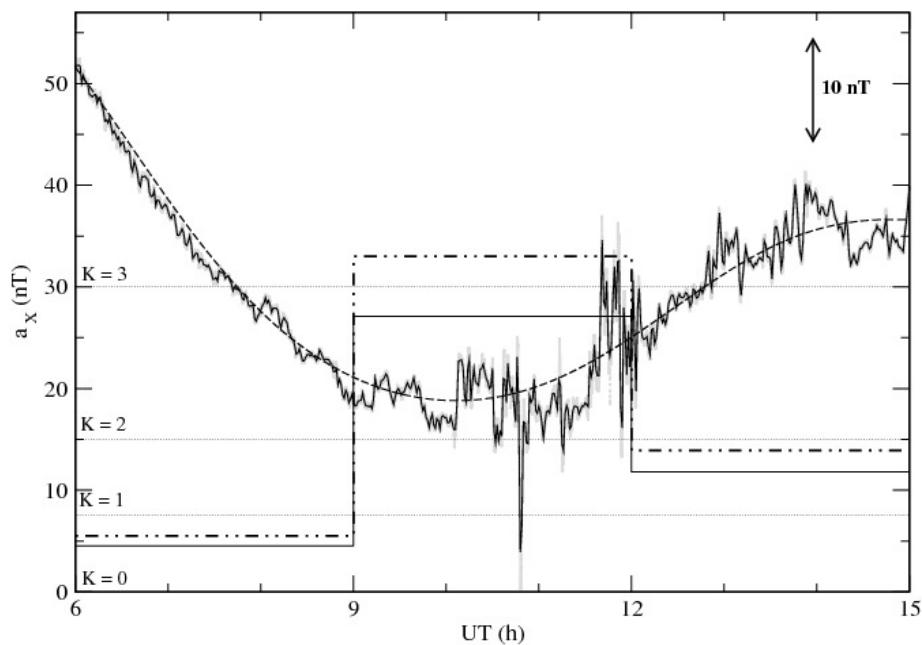

**Figure 3.** Magnetograms of the North component of the magnetic field at PAF (October 04$^{th}$, 2009) during three consecutive UT 3-hour intervals. The difference between the 1-minute values magnetogram (solid black curve) and the 1-second values one (solid grey curve) results in a systematic underestimation of the $a_X$ 3-hour ranges (1-minute values: solid step function; 1-second: dot-dashed step function). The up down double arrow indicates the scale, in nT, of the magnetograms.

In order to get quantitative information on the energy contained in the range 1/2 and 1/120 Hz between quiet and disturbed periods, the energy density spectrum *E(f)* was calculated for four different magnetic conditions (Figure 4): very quiet (*Am = 4* nT), quiet (*Am = 15* nT), moderately disturbed (*Am = 30* nT), and disturbed magnetic situations (*Am = 64* nT). The results make clear that the level of the energy in the 1/2 to 1/120 Hz frequency range is significant except during very quiet periods. This suggests that downward bias might occur for most of the 3-hour interval.

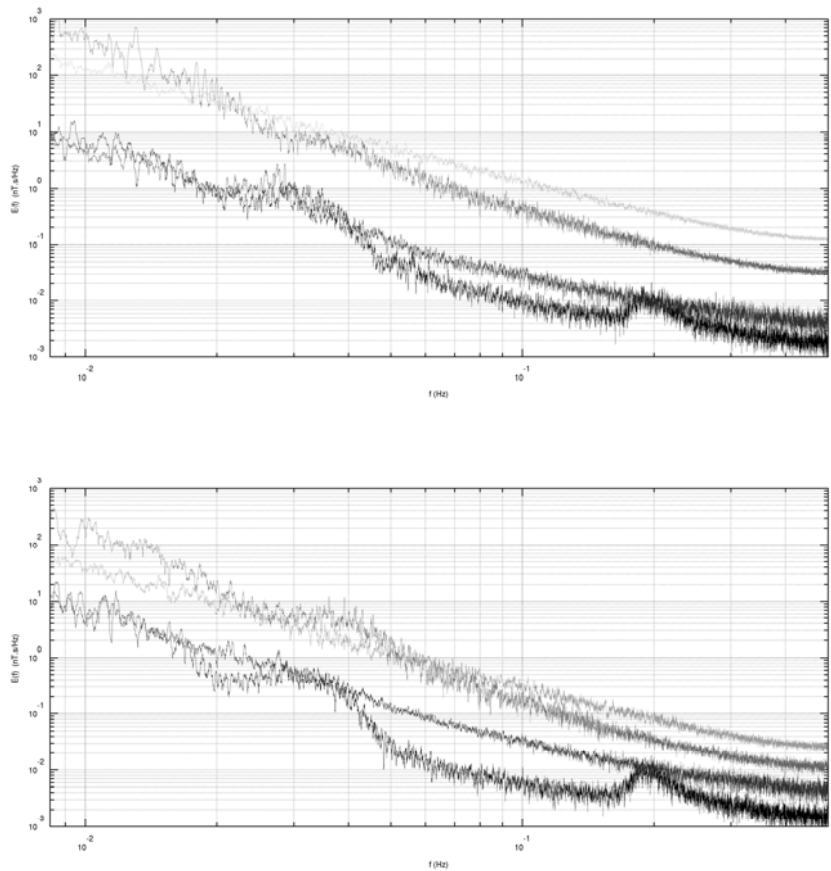

**Figure 4.** Energy Density Spectrum for four different magnetic conditions on the X (upper panel) and Y (lower panel) components of the magnetic field at PAF. From black upwards to the lighter grey curve: very quiet magnetic conditions (April 13$^{th}$, 2010; Am = 4 nT), quiet (March 10$^{th}$, 2010; Am = 15 nT; black curve); moderately disturbed (April 12$^{th}$, 2010; Am = 30 nT), and disturbed (April 5$^{th}$, 2010; Am = 64 nT)

## 4.1 Effect of the sampling interval on the K index

The comparison between $K_{60s}$ and $K_{1s}$ values shows that $K_{60s}$ is underestimated ($K_{60s}(t) = K_{1s}(t) - 1$) for 12% of the intervals. Figure 5 shows that the higher percentage of dissimilarity is observed for lower $K$ values: 98% of the intervals for which $K_{60s} \neq K_{1s}$ corresponds to $K_{60s} < 2$.

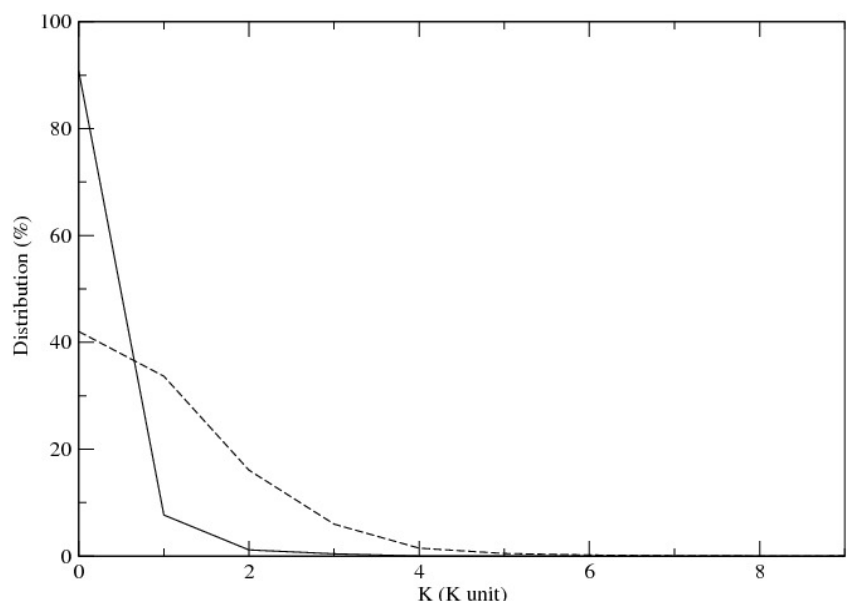

**Figure 5.** Distribution in percentage of the $K_{60s}$ values (computed using the 1-minute data) for those of the 3-hour intervals for which the $K_{60s}$ value is different from (and smaller than) the $K_{1s}$ one (690 intervals, solid curve) and for all the 3-hour intervals along the two years of data (5641 intervals, dashed curve)

The energy for frequencies higher than 1/300 Hz is essentially driven by magnetic pulsations. Even if magnetic pulsations may range from tenths of nT to tens of nT, the amplitudes of the related oscillations remain of the order of a few nT (especially during the selected period of low and moderate geomagnetic activity). Such amplitudes correspond to a fraction of the class of ranges that decreases with increasing $K$ values. This part is large for low $K$ values ($K=0$ or $K=1$) whereas it is very small for high $K$ values.

This argument concurs with that of Niblett et al. (1984) who explained that "*the lower K levels (…) contain much of the low energy pulsational activity with periods of 5 minutes or less*".

## 4.2 Diurnal and seasonal dependence of the sampling rate effect

Niblett et al. (1984) investigated for the OTT observatory the diurnal and seasonal distribution of the intervals for which $K_A$ and $K_R$ have different values. Their results illustrate the intervals for which $K_{60s} \neq K_{1s}$ shows a local time (LT) dependency. We made a similar study for PAF observatory, with $K_{1s}$ and $K_{60s}$. Figure 6 presents the diurnal LT distribution of discrepant $K$ for both observatories. For each station, a clear maximum appears around 1300 LT, just after the solar maximum irradiance. The secondary maximum observed in the OTT data is not observed in the PAF data.

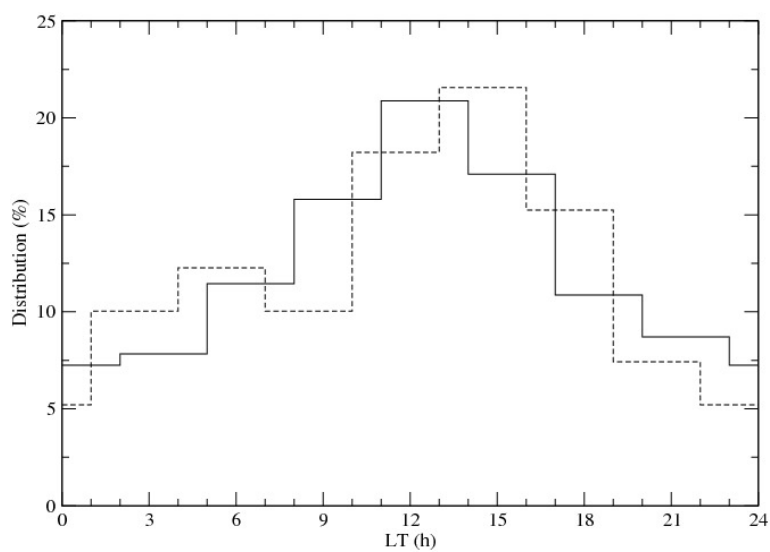

**Figure 6.** Diurnal distribution of intervals for discrepant K as a function of LT. The dashed curves correspond to $K_A$ and $K_R$ for OTT in 1982 (from Niblett et al., 1984; LT = UT - 5h); the solid curve corresponds to $K_{1s}$ and $K_{60s}$ for PAF in 2008.5-2010.5 (this study; LT = UT + 5h).

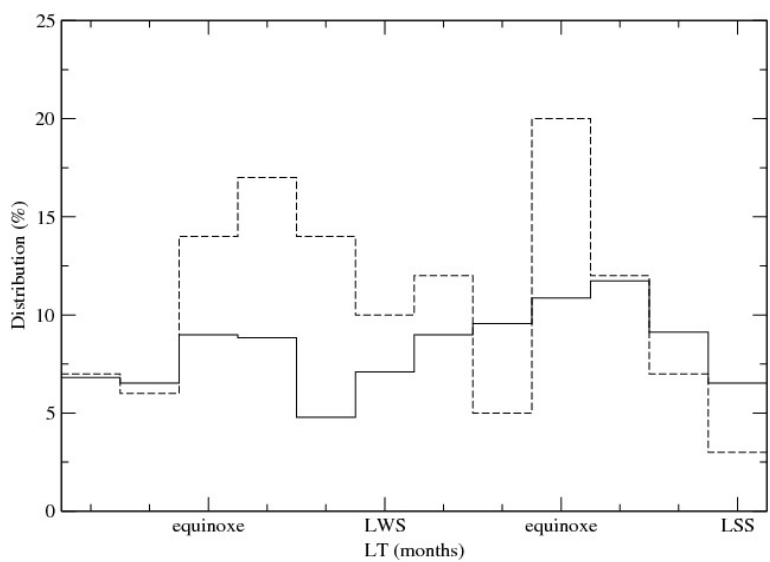

**Figure 7.** Monthly distribution of intervals of discrepant K as a function of the seasons. The dashed curves correspond to $K_A$ and $K_R$ for OTT in 1982 (from Niblett et al., 1984; LT + 6 months); the solid curve corresponds to $K_{1s}$ and $K_{60s}$ for PAF in 2008.5-2010.5 (this study). LWS means "Local Winter Solstice, and LSS means "Local Summer Solstice".

Figure 7 presents the seasonal distribution of discrepant *K* for both observatories. To take into account the latitudinal difference between the two stations, the histograms from Niblett et al. (1984) was shifted by 6 months.

The OTT distribution of discrepant *K* has a clear maximum around each equinox. The PAF distribution has not such a sharp maximum even if two smaller bumps, each centered on an equinox, may be noticed. Niblett et al. interpreted the OTT distribution as the consequence of the seasonal modulation of the pulsations. Since the intensity of the pulsations also varies jointly with the solar cycle, the seasonal modulation is expected to be weaker during the solar minimum than during the solar maximum, as observed in this study.

## 5   CONCLUSION

The present work shows that the $K_{60s}$ are biased downward with respect to $K_{1s}$ by one level in about 12% of the intervals. 98% of these low values occurred at the $K = 0, 1, 2$ levels. These results concur with those already obtained by Niblett et al. (1984), keeping in mind the large difference in geomagnetic activity level between the two considered datasets. Basically, the downward bias leads to an underestimation of the magnetic activity during magnetic quietness while it has very limited effect during a disturbed period.

On the one hand, the bias was observed when rapid geomagnetic fluctuations with periods less than about 200s occurred because these variations are severely attenuated in 1-minute data. A more precise determination of the periods responsible for the bias would require studying data series with various sampling intervals in the range of 1s to 60s.

On the other hand, the diurnal and seasonal distributions of discrepant *K* intervals do not act in synchrony with the solar cycle. The dependency remains to be investigated.